\documentclass[preprint2]{aastex}

\usepackage{amsmath}
\usepackage{graphicx}
\usepackage{natbib}
\usepackage{txfonts}
\usepackage{amssymb}
\usepackage{rotating}

\bibpunct{(}{)}{;}{a}{}{,}
\newcommand{\myarcsec}{\hbox{$.\!\!^{\prime\prime}$}}
\newcommand{\myarcmin}{\hbox{$.\!\!^{\prime}$}}
\newcommand{\ks}{\hbox{$K_{\rm s}$}}

\shorttitle{Gemini Frontier Fields: MACS J0416.1-2403 and A2744}
\shortauthors{Mischa Schirmer et al.}

\begin{document}
\bibliographystyle{aa}

\title{Gemini Frontier Fields: Wide-field Adaptive Optics $K_s$-band Imaging of\\
the Galaxy Clusters MACS J0416.1-2403 and Abell 2744\footnote{
  {\bf Based on observations} obtained at the Gemini Observatory, which is operated
  by the Association of Universities for 
  Research in Astronomy, Inc., under a cooperative agreement with the NSF 
  on behalf of the Gemini partnership: the National Science Foundation (United
  States), the Science and Technology Facilities Council (United Kingdom), the
  National Research Council (Canada), CONICYT (Chile), the Australian Research Council
  (Australia), Minist\'{e}rio da Ci\^{e}ncia, Tecnologia e Inova\c{c}\~{a}o (Brazil) 
  and Ministerio de Ciencia, Tecnolog\'{i}a e Innovaci\'{o}n Productiva (Argentina).
  {\bf Based on observations} made with ESO Telescopes at the La Silla and Paranal 
  Observatories, Chile.}
}


\author{M. Schirmer\altaffilmark{1}, 
  E. R. Carrasco\altaffilmark{1}, 
  P. Pessev\altaffilmark{1},
  V. Garrel\altaffilmark{1},
  C. Winge\altaffilmark{1}, 
  B. Neichel\altaffilmark{2,1},
  F. Vidal\altaffilmark{3,1}}
\affil{\altaffilmark{1}Gemini Observatory, Casilla 603, La Serena, Chile}
\affil{\altaffilmark{2}Aix Marseille Universit\'{e}, CNRS, Laboratoire d'Astrophysique de Marseille, 
  UMR 7326, 13388, Marseille, France}
\affil{\altaffilmark{3}Observatoire De Paris, Place Jules Janssen, 92190 Meudon, France}
\email{mschirme@gemini.edu}

\begin{abstract}
We have observed two of the six Frontier Fields galaxy clusters, MACS J0416.1-2403 and Abell 2744,
using the Gemini Multi-Conjugate Adaptive Optics System (GeMS) and the Gemini South Adaptive Optics Imager 
(GSAOI). With 0\farcs08$-$0\farcs10 FWHM our data are nearly diffraction-limited over a
$100^{\prime\prime}\times100^{\prime\prime}$ wide area. GeMS/GSAOI complements the \textit{Hubble Space Telescope} 
(\textit{HST}) redwards of $1.6\mu$m with twice the angular resolution. We reach a $5\sigma$ depth of 
$\ks\sim25.6$ mag (AB) for compact sources. In this paper we describe the observations, the data 
processing and the initial public data release. We provide fully calibrated, co-added images matching 
the native GSAOI pixel scale as well as the larger plate scales of the \textit{HST} release, adding to the 
legacy value of the Frontier Fields. Our work demonstrates that even for fields at high galactic latitude, 
where natural guide stars are rare, current multi-conjugated adaptive optics technology at 8m-telescopes has 
opened a new window on the distant Universe. Observations of a third Frontier Field, Abell 370, are planned.
\end{abstract}


\section{Introduction}
Strong gravitational lensing by massive clusters of galaxies provides us with magnified views of high redshift 
galaxies in the young Universe. The \textit{HST} Frontier Fields campaign observes six strong lensing clusters
in the $0.43-1.6\;\mu$m wavelength range, using the Advanced Camera for Surveys (\textit{ACS}) and the Wide 
Field Camera 3 (\textit{WFC3}). Complementary observations have been obtained with Chandra (PI: Murray, S.), 
Spitzer (PIs: Soifer, T., Capak, P.), Subaru \citep{pcf11}, the VLT \citep{orn11} and the AAT \citep{emb14}.

\begin{table*}
\caption{Summary of the observing nights as of February 2015}
\label{observations}
\begin{tabular}{lcccccc}
\noalign{\smallskip}
\hline
\hline
\noalign{\smallskip}
Target & Night (UT) & Exposures & $\langle {\rm AO\;seeing}\rangle$ & Nat. seeing\\
\hline
\noalign{\smallskip}

MACS J0416.1-2403 & 2014-01-13 & $40\times120$s & 0\farcs076 & 0\farcs4$-$0\farcs7\\
MACS J0416.1-2403 & 2014-01-14 & $37\times120$s & 0\farcs077 & 0\farcs6$-$0\farcs9\\
MACS J0416.1-2403 & 2014-01-19 & $34\times120$s & 0\farcs175 & 0\farcs9$-$1\farcs1\\
MACS J0416.1-2403 & 2014-01-20 & $47\times120$s & 0\farcs103 & 0\farcs5$-$0\farcs9\\
MACS J0416.1-2403 & 2014-01-22 & $16\times120$s & 0\farcs105 & 0\farcs5$-$0\farcs7\\
Abell 2744        & 2014-09-07 & $40\times120$s & 0\farcs110 & 0\farcs7$-$1\farcs2\\
\hline
\end{tabular}
\end{table*}

\textit{HST/WFC3} is sensitive to near-infrared radiation shorter than 
$1.7\mu$m\footnote{\tt http://www.stsci.edu/hst/nicmos/documents/handbooks/
current\_NEW/c04\_imaging.6.8.html}. This design was chosen (1) to allow for simpler thermo-electric 
cooling of the detector, and (2) because \textit{HST} itself contributes significantly to the thermal 
background and thus would offer only a small gain with respect to ground-based observatories at 
$\gtrsim2\mu$m. While the Earth's atmosphere has a suitable observing window between 
$2.0\mu$m and $2.3\mu$m (\ks-band), its turbulence limits the resolution and the depth with classical 
imaging. Using multi-conjugated adaptive optics (MCAO) \citep{rmv00,elr00}, these adverse effects
can be overcome for fields as large as one arcminute or more. In theory, diffraction limited 
observations are possible, given a sufficient number of bright natural guide stars (NGS) and a 
good laser return for the artificial laser guide stars (LGS). Nowadays, such wide-field diffraction 
limited observations can be obtained in the near-infrared at Gemini South with 
GeMS\footnote{\tt http://www.gemini.edu/sciops/instruments/gems/} \citep{rnb14,nrv14} and the
GSAOI\footnote{\tt http://www.gemini.edu/sciops/instruments/gsaoi/}
\citep{mhs04,cem12} camera.

GeMS is the first MCAO system in use at an 8m telescope. It delivers uniform and nearly diffraction-limited 
images over a $2^\prime$ field at near-infrared wavelengths. GeMS uses five sodium LGS 
and needs three NGS to compensate for tip-tilt and focus, reducing plate-dynamical errors
(plate scale variations). To achieve optimum performance, the NGS should be positioned as 
close as possible to an equilateral triangle around the science target. Best constellations 
(\textit{asterisms}) are the ones that cover most of the field. The larger the angular 
separations between the stars, then the lower the plate scale error will be over the 
imaged field. Unfortunately, this condition is usually not met at high galactic latitude where the
stellar density is low. With sub-optimal asterisms the PSF will become larger and non-uniform. Only one 
sufficiently bright NGS is available for each of our targets, yet we still achieve an image seeing of 
100 milli-arcseconds or better.

The focus in the remainder of this paper is on the observations (Section 2), and on various data 
reduction aspects such as background modeling and astrometry (Section 3). In Section 4 we 
summarize the properties of the coadded images and describe the data release.

\section{\label{sectionobs}Observations}
GSAOI is a near-infrared adaptive optics (AO) camera designed to work with GeMS. Its focal plane is 
formed by a $2\times2$ mosaic of Hawaii-2RG $2048\times2048$ pixel arrays with 2\farcs8$-$3\farcs0 
wide gaps. Images are recorded in a $85^{\prime\prime}\times85^{\prime\prime}$ field of view with a 
plate scale of 0\farcs0197 pixel$^{-1}$. The \ks-band filter in GSAOI has 50\% cut-on and cut-off 
wavelengths of $1.99\mu$m and $2.31\mu$m, respectively. More details about GSAOI can be found on the 
instrument web page.

The GSAOI pointings for MACS J0416.1-2403, Abell 2744 and Abell 370 are shown in the
Appendix (Figs. \ref{macs0416_layout}, \ref{a2744_layout} and \ref{a370_layout}, respectively). 
We have overlaid the currently available \textit{HST} data, as well as the areas of highest 
strong lensing magnification for $z=9$ sources as calculated by \cite{rjl14}. In case of
Abell 2744 we cannot cover the area of highest magnification due to the large angular separation
of the NGS from the cluster center.

MACS J0416.1-2403 was observed to the planned depth with GSAOI using director's discretionary time 
(Program ID: GS-2013B-DD-1). To optimize the observations, the NGS was located outside the GSAOI 
field of view, but within the patrol field area of GeMS. Due to a setup error (the NGS is a high 
proper motion star), the images observed during the first two nights were recorded with a South-Eastern 
offset of $24^{\prime\prime}$. As a result, the NGS appears in one of the four 
arrays. The images recorded during the other three nights have the correct base position. 

Regular $3\times3$ dither patterns (repeated several times) with a step size of $7^{\prime\prime}$
were used to cover the gaps between the arrays. Processing of the data has revealed that this strategy 
does not allow to fully suppress background residuals and cross-talk (Sect. \ref{datareduction}). 
Hence we switched to random dithers within a $14^{\prime\prime}$ wide box for Abell 2744 (and Abell 370), 
resulting in significantly improved background characteristics.

The observing log is presented in Table \ref{observations}. The targets, dates, number of images and 
exposure times are listed in columns 1, 2 and 3, respectively. For the remainder of the Abell 2477 
and Abell 370 data we will use 180s instead of 120s to reduce the overhead. Columns 4 and 5 show the 
average corrected AO FWHM within $1^{\prime}$ of the NGS, and the natural seeing recorded by the DIMM, 
respectively. 2014-01-19 was significantly worse with an average AO-corrected seeing of 0\farcs17. Data 
from that night was included in the low resolution stack, only (see Sect. \ref{release}). Observations
for Abell 2744 have been started in September 2014 (Program ID: GS-2014B-DD-1, 25\% complete by the 
time of writing). Abell 370 will follow in 2015.

\section{Data reduction}
\label{datareduction}
Data processing was done with {\tt THELI} \citep{sch13,esd05}, following the example in 
Appendix B of \cite{sch13}. Specific treatments for the GSAOI data are motivated below.

\subsection{Two-pass background modeling}
\label{backgroundmodelling}
To summarize, we have subtracted the median of 8 conterminous exposures to remove the background 
from the image sequences. This corresponds to a floating time window of 22 minutes duration. 
Windows of $15-30$ minutes length yield good background correction as well. Shorter windows (5 or less 
images) cause artifacts near extended objects (insufficient statistics due to masking) and increased 
noise. Longer windows (e.g. $35-45$ minutes) may undersample temporal variations in atmospheric emissivity, 
increasing the background residuals (see Sect. \ref{skysub}). Owing to its comparably small field of view, 
GSAOI is less sensitive to the latter effect than other near-infrared imagers.

The average background signal is $7490\pm440$ ADU or 12.58 mag arcsec$^{-2}$ (VEGA) in \ks-band.
This is close to the long-term average of 12.62 mag arcsec$^{-2}$, including a thermal contribution 
of $\sim0.7$ mag arcsec$^{-2}$ from the warm MCAO enclosure (see also Sect. \ref{skysub}).

In the first step of the two-pass background subtraction, the exposures were median combined without 
masking to get rid of the bulk of the sky signal. During the second step, {\tt THELI} 
applies \textit{SExtractor} \citep{bea96} to the corrected images from the first pass, 
computing mask images. Without these masks the fainter halos of the cluster galaxies bias the 
background model toward higher values, causing dark halos in the coadded image. 

\begin{figure}[t]
  \includegraphics[width=1.0\hsize]{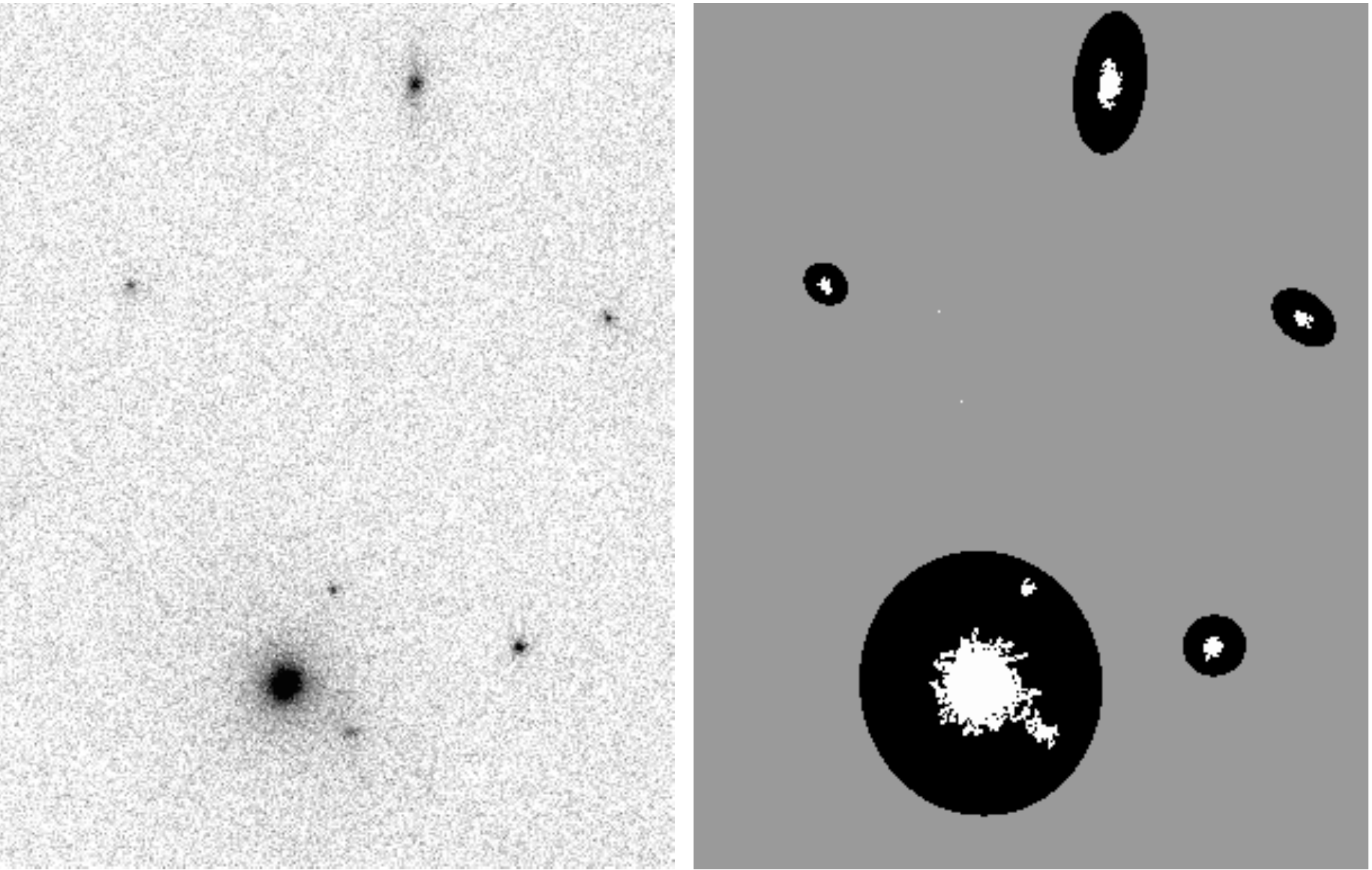}
  \caption{\label{maskexpand}{Expansion of object masks to avoid biasing of the background
      models by faint galaxy halos. Left: Part of a GSAOI image after first pass 
      background modeling. Right: \textit{SExtractor} mask before (white) and after 
      (black) expansion with $R=8$, using {\tt DETECT\_THRESH$\,=\,$1.2} and 
      {\tt DETECT\_MINAREA$\,=\,$20}.}}
\end{figure}

We have found that the \textit{SExtractor} masks are too small. A module was added to {\tt THELI} 
that optionally enlarges the masks. It uses an alternative parametrization of the best-fit 
ellipse to an object's isophotes. A pixel with image coordinates $x$ and $y$ is located inside the 
ellipse if
\begin{equation}
{\rm C_{XX}} (x-\bar{x})^2 + {\rm C_{YY}}(y-\bar{y})^2 + {\rm C_{XY}} (x-\bar{x})(y-\bar{y}) <= R^2\;,
\label{maskexpansion}
\end{equation}
where $\bar{x}$ and $\bar{y}$ represent the object's centroid (first moment), and $\rm C_{XX}$, 
$\rm C_{YY}$ and $\rm C_{XY}$ are \textit{SExtractor} parameters calculated from the second 
brightness moments. A choice of $R\sim3$ reproduces the outer isophote of a 
detection\footnote{For details see the \textit{SExtractor} user manual v2.13, Sect. 10.1.6)}.
Only for $R=8$ or higher the over-correction of the sky disappeared in the co-added images. 
This \textit{mask expansion} is available in {\tt THELI} as of v2.9.0, with $R$ being 
referred to as the \textit{mask expansion factor}. It can be selected during background 
modeling, collapse correction (see below), and individual sky subtraction. Figure 
\ref{maskexpand} illustrates the effect.

Note that low surface brightness features larger than the dither box (such as intra-cluster light)
escape this masking process and are suppressed in the co-added images. This can be avoided by 
interspersing blank sky fields into the observing sequence.

\begin{figure*}[t]
  \includegraphics[width=1.0\hsize]{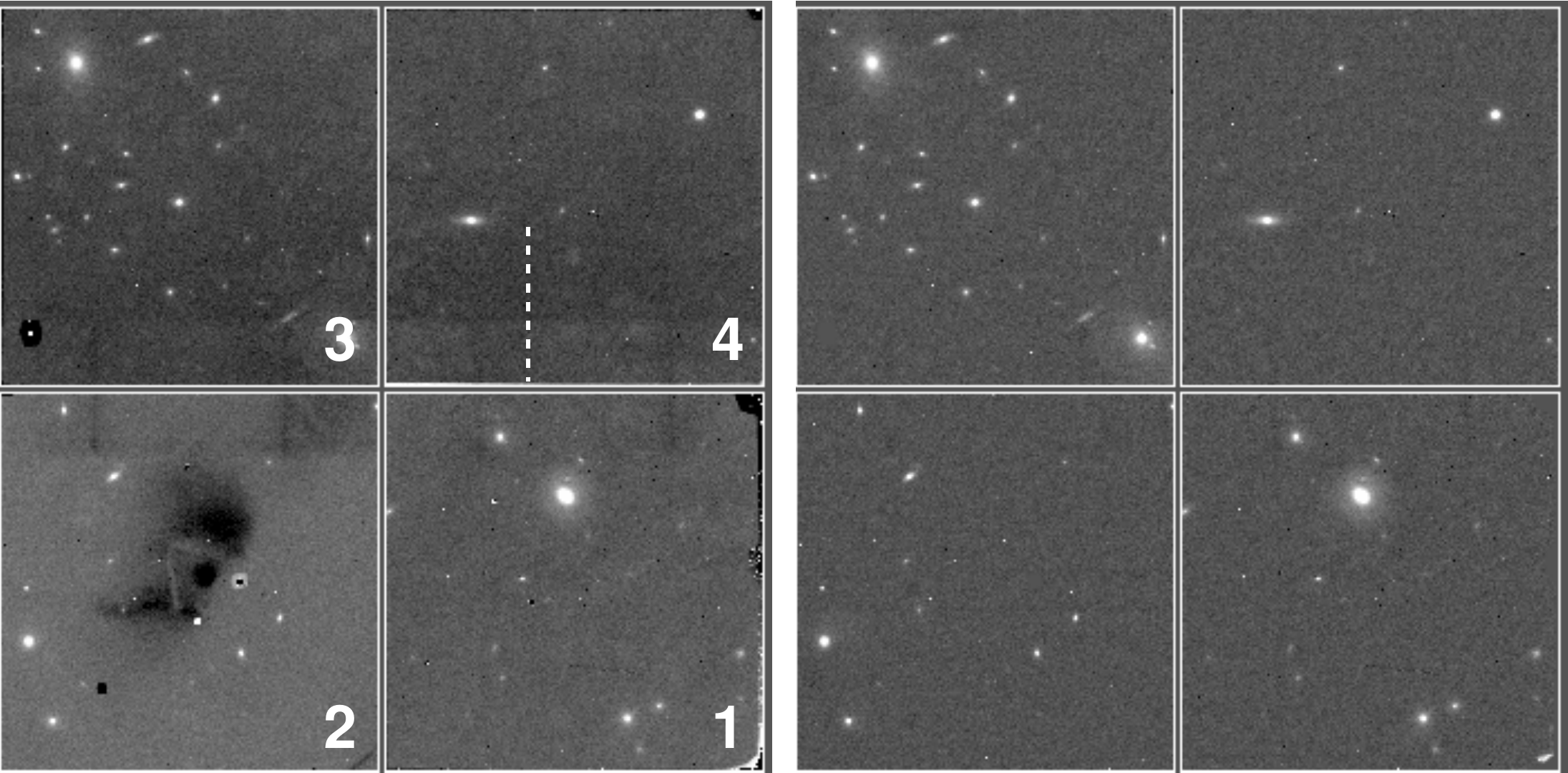}
  \caption{\label{mosaic}{Left: The four GSAOI detector arrays (labeled). A reset anomaly (dark 
      extended spot) appears in array \#2 at the beginning of each observing sequence. Independently, 
      horizontal striping with a low amplitude occurs in $\gtrsim80$\% of exposures, affecting all 
      four detectors. We illustrate the latter by plotting the mean row values along the dashed 
      white line in array \#4 both before and after correction (see Fig. \ref{vcut}).
      Right: Same exposure after correction for reset anomaly and striping. Note that this is a 
      high contrast display of $12\times12$ binned data, exaggerating the effects.}}
\end{figure*}

\subsection{Reset anomaly and background gradients}
GSAOI shows an unstable reset anomaly in array \#2 for its medium and faint readout modes, probably 
due to a bias/dark problem in the detector controller. Every first image in an exposure series is 
affected by it, as well as the image taken immediately after an interruption (e.g. because of the 
laser being shuttered for air planes or satellites). The anomaly is evident in the data after background 
subtraction. Figure \ref{mosaic} shows that we can correct it by subtracting a rescaled median model of 
the affected images. The rescaling factors (between 0.2 and 1.2) were determined manually.

\begin{figure}[t]
  \includegraphics[width=1.0\hsize]{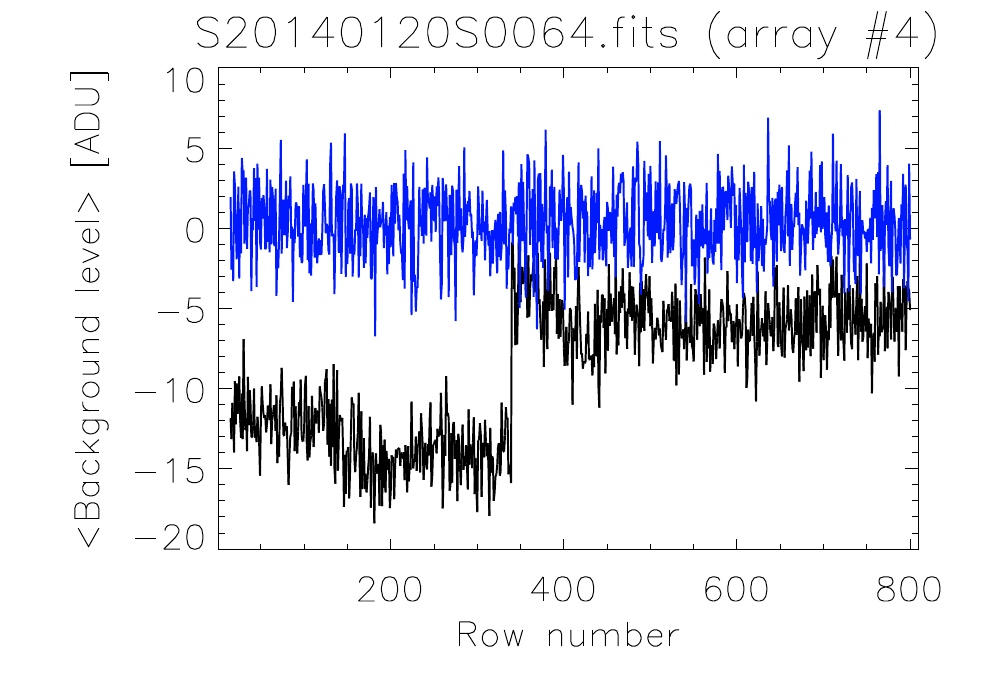}
  \caption{\label{vcut}{Mean row values before (black line) and after (blue line) correction
      for striping. The data was extracted along the dashed line shown in Fig. \ref{mosaic}.}}
\end{figure}

In addition to the reset anomaly, and independent of it, horizontal stripes are found in 84\% 
of the MACS J0416.1-2403 data (and to a lesser degree for Abell 2744). They run parallel to the 
interface between arrays \#2/3 and \#1/4 (see Figs. \ref{mosaic} and \ref{vcut}). Both the amplitude 
and the angular extent of the stripes vary between exposures. They affect an identical number 
of rows in all four arrays. The amplitude is much lower than the reset anomaly, in the worst case
reaching 0.2\% of the background level ($3-4$ times smaller than the sky noise). For deep coadded 
images this effect must be corrected for by subtracting an average column profile 
(\textit{collapse correction} in {\tt THELI}). The calculation of the profile includes
object masking (see Sect. \ref{backgroundmodelling}).

\subsection{\label{skysub}Sky subtraction}
Background modeling with a floating median (Sect. \ref{backgroundmodelling}) may leave residuals
in the data due to undersampled temporal variations of the sky. We correct individual exposures by 
masking all objects (Sect. \ref{backgroundmodelling}), using a conservative mask expansion factor of 
$R=20$. The masked images are convolved with a Gaussian and subtracted from their originals. For 
MACS J0416.1-2403, best results in terms of preserving faint extended galaxy halos while minimizing 
background inhomogeneities were obtained with a $3^{\prime\prime}$ wide kernel. Abell 
2744 required a $6^{\prime\prime}$ kernel due to a few low surface brightness galaxies 
which were undetected (and thus unmasked) in individual exposures.

The coadded images (normalized to 1s exposure time) have a mean background level of zero.
Without sky subtraction the normalized background value would be $62\pm4$ ADU. Effective gains 
are listed in the FITS headers. Note that due to the dithering of mosaiced data the total exposure 
time varies across the image. Local noise levels must be estimated using the coadded weight 
maps.

\subsection{Image persistence and crosstalk}
Saturated objects leave a weak imprint in subsequent GSAOI exposures (image 
persistence). In the case of MACS J0416.1-2403, the only saturating source is a bright field 
star (Fig. \ref{macs0416_layout}), leaving a charge residual of $\sim0.2\%$. We have masked 
these ghosts manually in the weight images as they did not average out entirely due to
the repeated dither pattern.

\begin{figure}[t]
  \includegraphics[width=1.0\hsize]{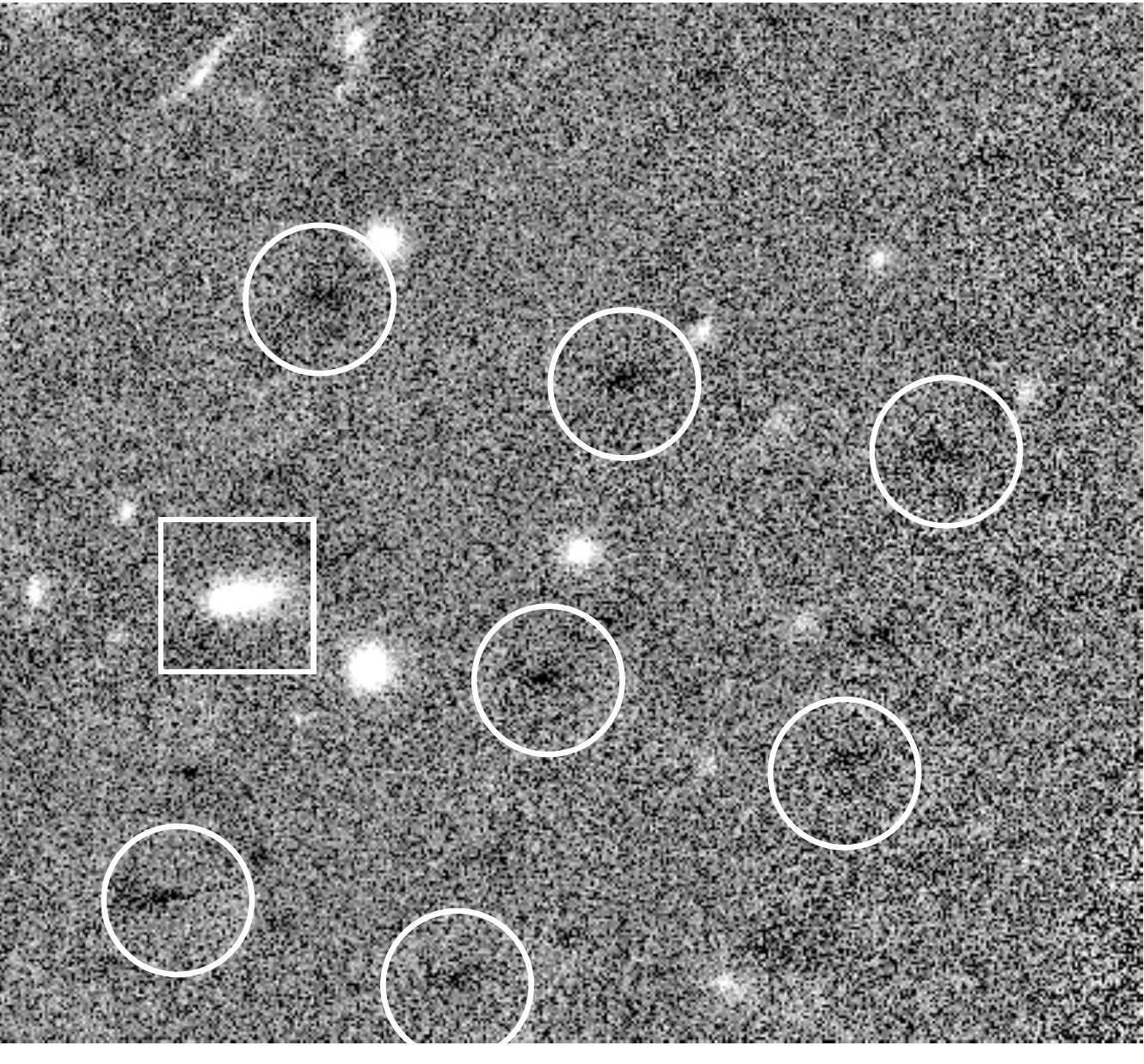}
  \caption{\label{xtalk_low}{A low-level memory effect (white circles) appears in certain 
      areas of the GSAOI detectors. In this example it is caused by the galaxy marked with 
      a white square. A neighboring object of similar brightness does not trigger the effect.}}
\end{figure}

The same star, located in array \#1, also causes significant negative crosstalk in one of the four 
output channels of array \#1. Contrary to image persistence, the position of this crosstalk is 
fixed with respect to the star and thus does not average out. We masked it manually in the weight maps.

Other memory effects exist in some GSAOI arrays. This is particularly noticeable in the low resolution 
(0\farcs06 pixel$^{-1}$) stack of MACS J0416.1-2403 with its high S/N ratio because of the effective 
$3\times3$ re-binning. Darker spots with a maximum amplitude of twice the background noise level 
(Fig. \ref{xtalk_low}) mirror the dither pattern. Only some parts of the arrays are affected, as 
neighboring sources of similar brightness do not cause ghosting. The origin of this behavior is still 
under investigation. It is of minor concern as it rarely coincides with images of astrophysical objects.
The non-repetitive dither pattern used for Abell 2744 (and Abell 370) fully suppresses this feature.

\begin{figure*}[t]
  \includegraphics[width=1.0\hsize]{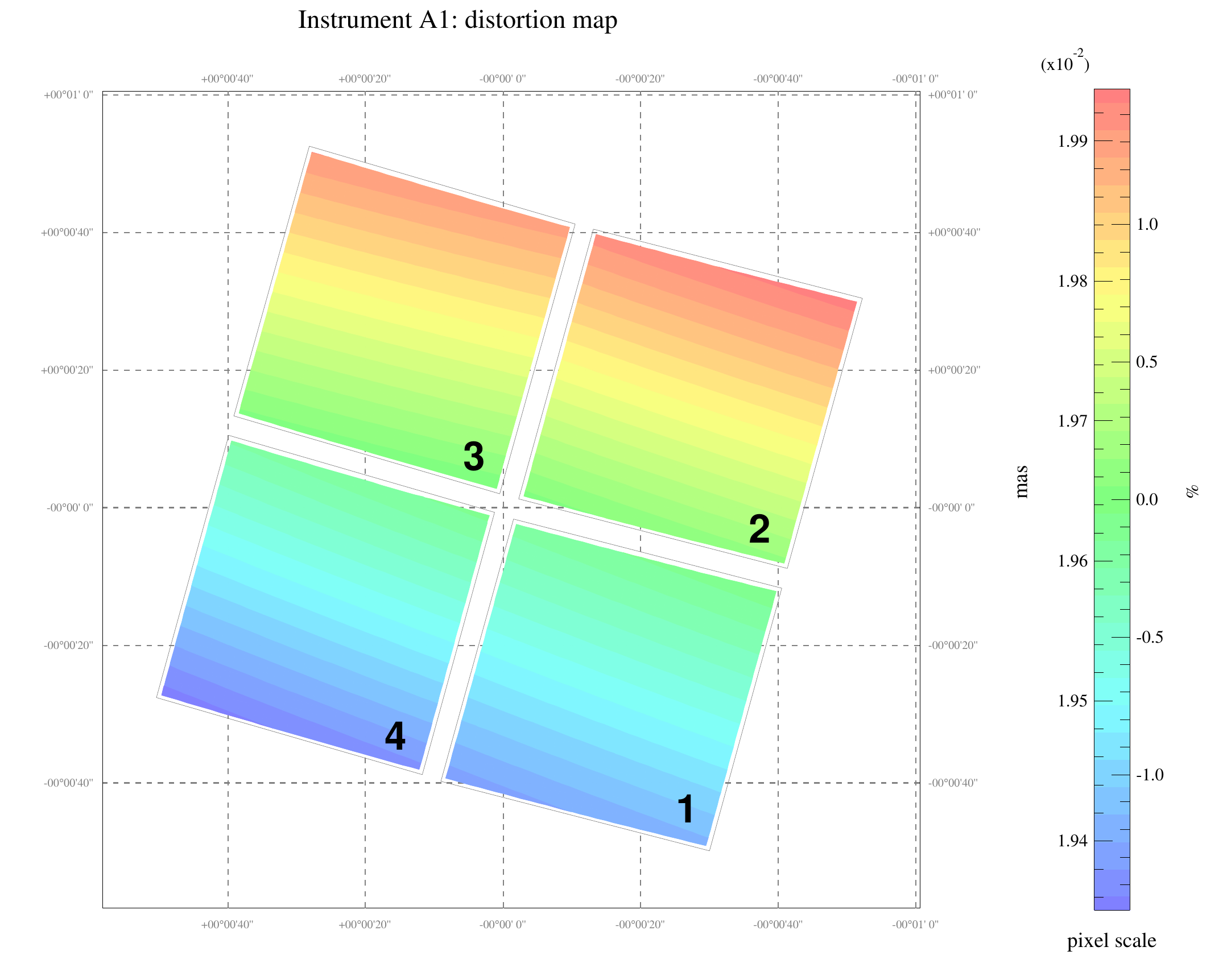}
  \caption{\label{gsaoi_dist}{Typical GSAOI distortion model. Note that \textit{Scamp} calculates 
      plate scale variations independently for each of the four arrays. The distortion is recovered 
      reliably from the dithered data despite the low source density.}}
\end{figure*}

\begin{figure*}[t]
  \includegraphics[width=1.0\hsize]{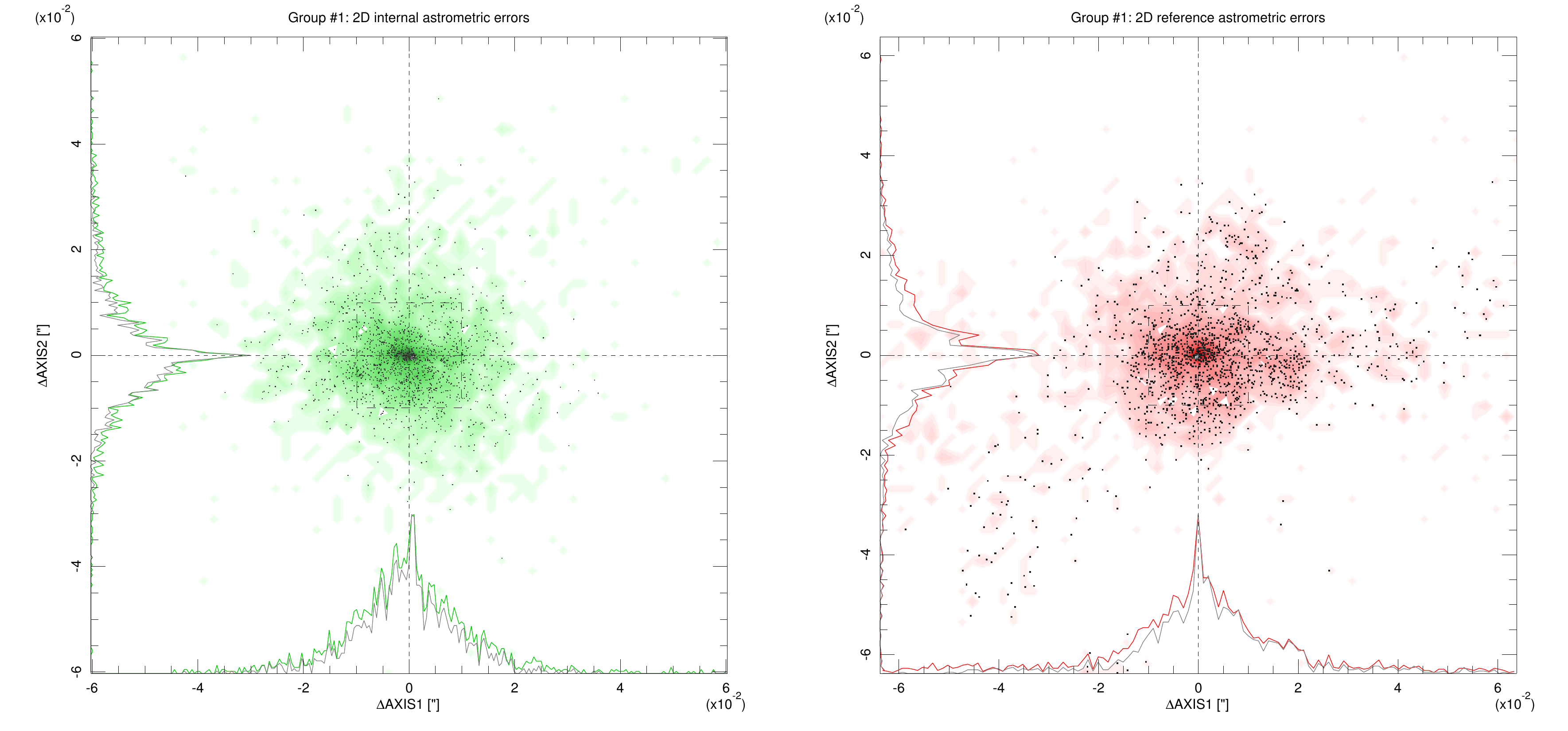}
  \caption{\label{gsaoi_astres}{Shown are the internal astrometric residuals (i.e. 
      between exposures, left panel) and the external residuals (with respect to the \textit{VLT/HAWK-I} 
      \ks-band image; right panel). Sub-pixel accuracy is achieved in both cases, despite only 
      a few dozen low-S/N detections of extended sources have been available for the 
      calibration.}}
\end{figure*}

\subsection{Astrometric calibration with Scamp}
{\tt THELI} uses \textit{Scamp} \citep{ber06} for the astrometric calibration, and the discussion below is 
specific in this context. Nonetheless, our results should provide a useful reference point when solving 
GSAOI astrometry with other software.

\subsubsection{Input catalogs and Scamp settings}
The setup error described in Sect. \ref{sectionobs} resulted in an offset of $24^{\prime\prime}$, which 
is not reflected in the headers of the archival data taken on 2014-01-13/14. The {\tt CRVAL1/2} header 
keywords were manually corrected to within $\sim 1^{\prime\prime}$, i.e. the typical WCS uncertainty
in raw GSAOI data. We then loaded \textit{SExtractor} source 
catalogs into \textit{Scamp} (v2.0.1), using a low detection threshold per pixel 
({\tt DETECT\_THRESH$\,=\,$1.2}) to maximize the number of usable sources. A minimum of 
{\tt DETECT\_MINAREA$\,=\,$20} pixels above this threshold was required to form a valid object (rejecting 
spurious sources, and because GSAOI data are oversampled). This resulted in an average of $11\pm5$ 
($8\pm3$) detections (mostly galaxies) per array for MACS J0416.1-2403 (Abell 2744). \textit{Scamp} uses 
its own internal S/N thresholds for object filtering. A very low setting is required to retain a sufficient 
number of objects for successful matching with the reference catalog. The final astrometric solution was 
obtained using a refined median estimate of the relative chip positions and orientations from all exposures 
in a night ({\tt MOSAIC\_TYPE$\,=\,$FIX\_FOCALPLANE}), and a 2nd order distortion model. We reproduce all 
relevant parameter settings in Table \ref{scampsettings}. A typical distortion model is shown in Fig. 
\ref{gsaoi_dist}, revealing a plate scale variation of 3\%.

\subsubsection{The need for secondary reference catalogs}
Astrometry with GSAOI is non-trivial. First, there is a large discrepancy between 
the angular resolutions of GSAOI and common all-sky astrometric reference catalogs. A single reference 
source (stellar or non-stellar) may be resolved into several components by GSAOI. Second, the AO 
field of view is comparatively small and at high galactic latitude no (or only a few) reference sources 
are available. Third, GSAOI works in the near-infrared, whereas reference catalogs are mostly based on 
optical data. This often results in very low densities of the 
reference catalog in highly reddened areas (2MASS is an exception, but it is comparatively shallow). 
In our case with mostly extragalactic sources, bluer reference catalogs are susceptible to sub-structures 
in galaxies whereas the redder science data mainly detect the galactic bulges and nuclei. Forth, GeMS 
introduces a significant field distortion by means of its optical relay (consisting of two off-axis parabolic 
mirrors). Fortunately, this distortion is static and, as we show below, it can be measured reliably even 
for sparse data. 

Lastly, the distortion correction is imperfect as the wavefront correction is a function of 
the atmospheric turbulence profile and e.g. the brightness of the LGS. The latter depends on the laser 
performance, the alignment of the laser beam transfer optics, and also the atmosphere's sodium layer (which 
has a strong seasonal dependence and is replenished by meteor showers). This affects mostly investigations 
where high precision astrometry is required. For more details see e.g. \cite{cbk09,abg13,nlr14}.

For our purposes, the key to successful WCS matching and sufficiently good distortion modeling 
are deeper secondary (or even tertiary) reference catalogs \citep[see also][]{sch13}, ideally constructed 
from observations at similar wavelengths.

\subsubsection{External astrometry}
For both galaxy clusters we tried different reference catalogs constructed from (1) the \textit{HST/ACS} 
F814W images (the \textit{HST/WFC3} near-infrared data do not fully overlap with the GSAOI pointings), 
and (2) deep \textit{VLT/HAWK-I} \ks-band observations (PI: Brammer, ESO program ID: 092.A-0472; image 
seeing 0\farcs29$-$0\farcs37). In case of MACS J0416.1-2403 we also tested a $R$-band image taken with 
the Wide Field Imager (\textit{WFI}) at the 2.2m MPG/ESO telescope \citep[][ESO program ID: 083.A-9026; 
image seeing 0\farcs95]{gsb14}. The co-added ground-based data were created with {\tt THELI} and calibrated 
against 2MASS \citep{scs06}.

Matching was successful for all three reference catalogs. However, calibration against the \textit{WFI} data 
caused large residuals ($30-70$ milli-arcseconds) and we do not consider it further. \textit{HST/ACS} and 
\textit{VLT/HAWK-I} catalogs have given similar results for MACS J0416.1-2403. For Abell 2744 we could not 
obtain a good distortion correction for array \#2 with the \textit{HST/ACS} catalog.

Particularly noteworthy is that the \textit{VLT/HAWK-I} catalog resulted in external residuals of $12-14$ 
milli-arcseconds, whereas for \textit{HST/ACS} we got $6-9$ milli-arcseconds. This is not surprising as the 
HST data have smaller measurement errors for the centroids due to the high angular resolution. 
However, the nominal \textit{Scamp} $\chi^2$ values for the \textit{VLT/HAWK-I} fit have been about $1-6$ per 
image for both clusters, whereas for \textit{HST/ACS} we have found $\chi^2=10-30$ for MACS J0416.1-2403, and 
$\chi^2=30-100$ for Abell 2744. The \textit{HST} data are significantly bluer ($0.8\mu$m) than the \textit{VLT} 
images ($2.2\mu$m) and thus more susceptible to sub-structures in the galaxies. The effect is immediately visible 
when over-plotting both reference catalogs over the GSAOI exposures. The centroids measured in the \textit{VLT/HAWK-I} 
data, taken in the same filter as the GSAOI images, align better for a larger number of objects. The higher 
$\chi^2$ for Abell 2744 is likely the result of a larger fraction of late-type galaxies (spirals and edge-on disks). 
The GSAOI pointing is further away from the cluster center than for MACS J0416.1-2403 (compare Figs. 
\ref{macs0416_layout} and \ref{a2744_layout}), and thus the effect of sub-structure is perhaps more pronounced.

We base our final external WCS calibration on the \textit{VLT/HAWK-I} data, which in turn have been calibrated 
against 2MASS. Note that in v1.0 of the \textit{HST} data systematic WCS offsets of 0\farcs17 and 0\farcs40 are 
present for MACS J0416.1-2403 and Abell 2744, respectively. Corrections 
for the \textit{HST} headers are as follows:
\begin{eqnarray}
\Delta{\rm{\tt CRVAL1}}&=&+5.0\times10^{-5} \;{\rm (MACS\; J0416.1)}\\
\Delta{\rm{\tt CRVAL2}}&=&-0.6\times10^{-5} \;{\rm (MACS\; J0416.1)}\\
\Delta{\rm{\tt CRVAL1}}&=&-5.8\times10^{-5} \;{\rm (Abell\; 2744)}\\
\Delta{\rm{\tt CRVAL2}}&=&+9.4\times10^{-5} \;{\rm (Abell\; 2744)}
\end{eqnarray}

\subsubsection{Internal astrometry}
The $\chi^2$ did not improve when switching from a 2nd to a 3rd order distortion polynomial. The higher 
order solution is poorly constrained, given the small number of sources, and significant discontinuities 
in the pixel scale occurred across array boundaries. Thus we have selected the simpler 2nd order hypothesis. 
The internal and external residuals for MACS J0416.1-2403 are shown in Fig. \ref{gsaoi_astres} and measure 
10 and 13 milli-arcseconds, respectively (same for Abell 2744). This is equivalent to half a GSAOI pixel 
and should be sufficient for most purposes of these particular data sets. The dominating factors limiting 
the fits are the low source densities and the large measurement uncertainties of the source centroids 
($0.1-0.3$ pixels or $2-6$ milli-arcseconds for galaxies). 

Astrometric solutions have been obtained on a nightly basis. We have also tested a common distortion model 
for all nights, and find it to be inferior. This is because the large angular offsets between the base 
positions of different nights introduce different distortion patterns as the NGS and LGS configurations 
move with respect to each other \citep{nlr14,nrv14,rnb14}. For another test we have split the nightly 
sequences into blocks of 30 minutes to sample a variable distortion component caused by flexure due to a 
changing gravity vector \citep{nlr14}. This effect becomes important when high precision relative astrometry 
(on the order of one milli-arcsecond and below) is required across the field \citep{nlr14,nrv14}. We 
find that for blocks with good AO-corrected seeing (80 milli-arcseconds) the internal residuals 
decrease from $10$ to $6-8$ milli-arcseconds, whereas for other blocks the quality of the fit was 
unchanged or worse. We thus decided to use the more stable nightly fits instead.

Better performances are hard to reach with GeMS at high galactic latitude with few low S/N detections 
per array, almost all of which extended, and just a single NGS. In comparison, for dense stellar fields 
such as NGC 1851 with $500-1000$ sources per array with ${\rm S/N}>10-100$, optimal NGS asterisms and 
in the absence of dithering, GeMS delivers relative astrometric accuracies of 0.4 milli-arcseconds and 
better \citep{nlr14,abg13,rnb12}. 

\begin{table}[t]
\caption{Useful \textit{Scamp} parameter ranges for our GSAOI data (see the \textit{Scamp} manual for details)}
\label{scampsettings}
\begin{tabular}{lr}
\noalign{\smallskip}
\hline
\hline
\noalign{\smallskip}
Parameter & Setting\\
\hline
\noalign{\smallskip}
{\tt POSANGLE\_MAXERR} & 0.5\\
{\tt POSITION\_MAXERR} & 0\myarcmin035\\
{\tt PIXSCALE\_MAXERR} & 1.03\\
{\tt DISTORT\_DEGREE} & 2\\
{\tt ASTREF\_WEIGHT} & 1\dots10\\
{\tt SN\_THRESHOLDS} & 2\dots4, 10\dots20\\
{\tt STABILITY\_TYPE} & {\tt INSTRUMENT}\\
{\tt MOSAIC\_TYPE} & {\tt FIX\_FOCALPLANE}\\
{\tt CROSSID\_RADIUS} & 0\farcs1\dots0\farcs6\\
{\tt MATCH} & {\tt Y}\\
\hline
\end{tabular}
\end{table}

\subsection{\label{photcal}Photometric calibration}
{\tt THELI} normalizes the gains of multi-chip cameras to the array with the lowest gain, 
i.e. array \#2 in case of GSAOI. The co-added images are scaled in ADU$\,$s$^{-1}$.
Transformation from VEGA to AB magnitudes in the \ks-band filter is obtained by 
adding 1.85 mag.

\subsubsection{Transparency}
\begin{figure}[t]
  \includegraphics[width=1.0\hsize]{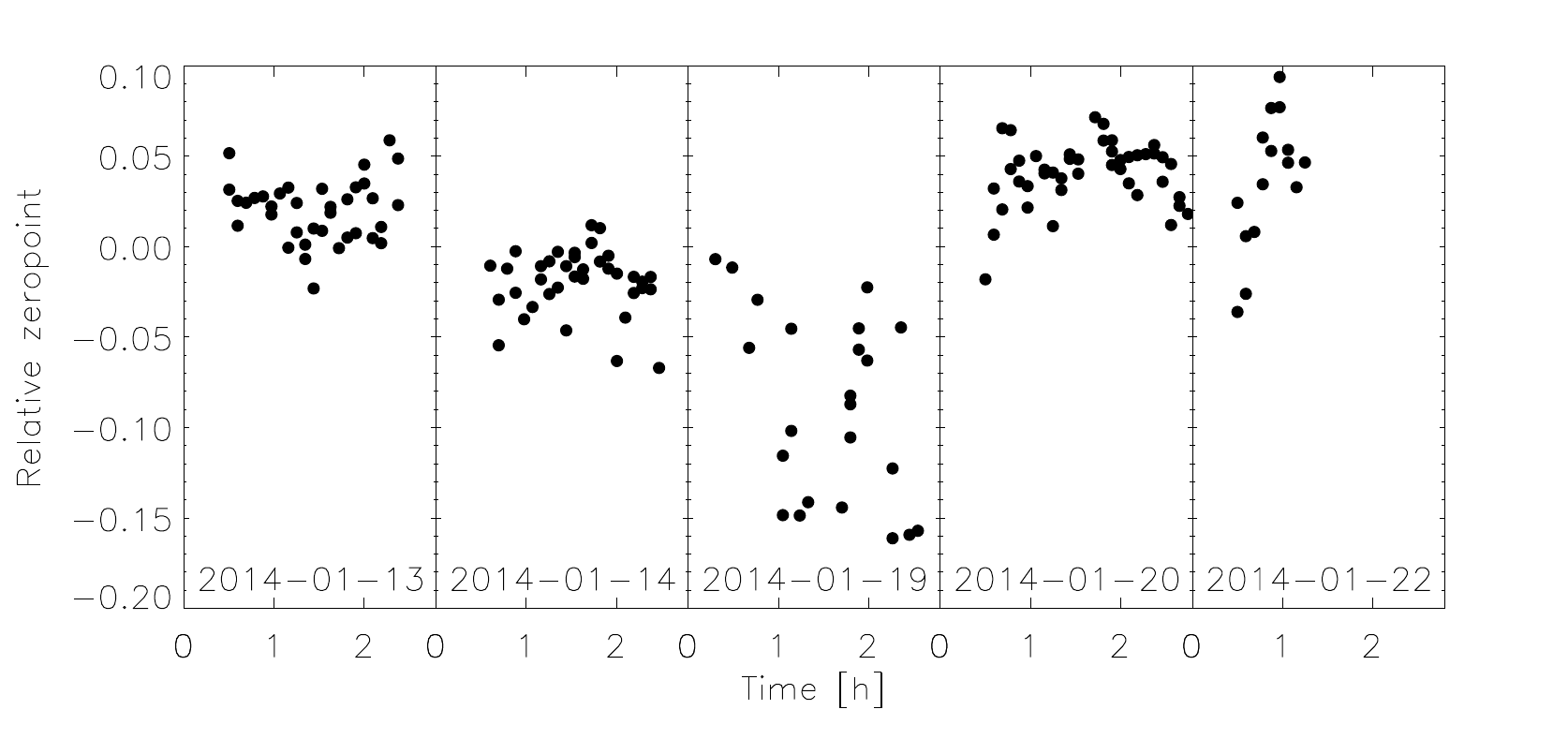}
  \caption{\label{frontier_rzp}{MACS J0416.1-2403: Variation of the relative photometric 
      zero-points between exposures.}}
\end{figure}

Sky transparency was good, yet somewhat unstable during all observing nights
(see Fig. \ref{frontier_rzp}). The instabilities were caused by a high inversion layer 
trapping significant amounts of humidity above the observatory. Such conditions may be difficult 
to recognize at night by observers. Other factors that contribute to the scattering between 
exposures are errors in the gain determination of the four arrays of $0.5\%-1.0\%$, and small 
differences in their quantum efficiencies. In \cite{cem12} we have observed similar uncertainties 
for various stellar and non-stellar fields.

\begin{table*}
\caption{Summary of the released co-added data. The FWHM is the mean seeing measured within 
$1^{\prime}$ of the NGS. Note that the Abell 2744 data will be incremented as by the time of 
writing only 25\% of the planned data have been taken.}
\label{datarelease}
\begin{tabular}{lccrcc}
\noalign{\smallskip}
\hline
\hline
\noalign{\smallskip}
Stack name & Filter & Pixel scale & $t_{\rm exp}$ [s] & FWHM\\
\hline
\noalign{\smallskip}
{\tt MACS\_J0416.1-2403\_GSAOI\_0.02\_deep.fits}   & \ks & 0\farcs02 & 15720 & 0\farcs085\\
{\tt MACS\_J0416.1-2403\_GSAOI\_0.02\_seeing.fits} & \ks & 0\farcs02 &  6600 & 0\farcs077\\
{\tt MACS\_J0416.1-2403\_GSAOI\_0.03\_deep.fits}   & \ks & 0\farcs03 & 15720 & 0\farcs086\\
{\tt MACS\_J0416.1-2403\_GSAOI\_0.03\_seeing.fits} & \ks & 0\farcs03 &  6600 & 0\farcs080\\
{\tt MACS\_J0416.1-2403\_GSAOI\_0.06\_deep.fits}   & \ks & 0\farcs06 & 18600 & 0\farcs095\\
\hline
{\tt A2744\_GSAOI\_0.02\_v0.1.fits} & \ks & 0\farcs02 & 4800 & 0\farcs100\\
{\tt A2744\_GSAOI\_0.03\_v0.1.fits} & \ks & 0\farcs03 & 4800 & 0\farcs105\\
{\tt A2744\_GSAOI\_0.06\_v0.1.fits} & \ks & 0\farcs06 & 4800 & 0\farcs120\\
\hline
\end{tabular}
\end{table*}

\subsubsection{Absolute zero-point}
For the absolute photometric calibration (MACS J0416.1-2403) two standards from the MKO catalog 
\citep{lcv06} were observed on 2014-01-13 at airmasses 1.13 and 1.35, 1h and 4h after the last 
science exposure was taken. Assuming that the mean transparency has not changed during the night,
we find a photometric zero-point of ${\rm ZP}=25.62\pm0.06$ mag (VEGA). This value is based on a 
linear two parameter fit for the zero-point and the extinction coefficient. We have neglected a small
color term ($\sim0.02$ mag) between the GSAOI \ks-band filter and the $K$-band filter used for the 
MKO system \citep{lcv06}. The uncertainty of $0.06$ mag includes the nominal uncertainty of the fit, 
contributions from the slightly non-photometric conditions and the uncertainties of the detector 
characteristics. For comparison, the only non-saturated star in the area detected by 2MASS has a 
\ks-band magnitude of $15.36\pm0.19$ mag (error from 2MASS photometry) after transformation to the 
MKO system, compared to 15.42 mag measured in our data.

To verify this calibration, we have re-observed MACS 
J0416.1-2403, Abell 2744 and a series of standard stars in \ks-band in a photometric night using 
the Flamingos-2 near-infrared imager at Gemini South. This independent Flamingos-2 calibration
falls within $0.002\pm0.02$ mag of the 2MASS field photometry. Aperture photometry of common 
sources in the combined Flamingos-2 and GSAOI data did not reveal an offset.

\subsubsection{Limiting magnitude}
The depth of the co-added images vary across the field due to the dithered 
multi-chip data and can be assessed by means of the released coadded weight maps. Another factor 
that becomes important (compared to classical imaging) is that most of the accessible faint field 
galaxies in these data have intrinsic sizes smaller than the natural seeing disk (0\myarcsec5-0\myarcsec7), 
but larger than the AO corrected seeing (0\myarcsec08). Therefore, a compact galaxy will benefit
more from the AO correction than an equally bright but more extended galaxy. This dependence of the 
field depth as a function of source FWHM is difficult to model due to the field galaxies' mostly
irregular morphologies, and we ignore it hereafter.

We cannot estimate the depth based on field stars as only a few point sources 
are found near MACS J0416.1-2403, none of which are near the detection limit. Instead, we use compact 
(yet extended) galaxies with half-light radii $\leq$0\farcs15 in the deep medium resolution stack (see 
Sect. \ref{release}). Based on the distribution of magnitude errors, we measure a $10\sigma$ detection 
limit of $\ks=22.4\pm0.2$ mag (VEGA) and extrapolate a $5\sigma$ limit of $23.8\pm0.4$ mag ($25.6\pm0.4$ 
AB mag). The depth of the intermediate Abell 2744 data has not yet been determined as only 25\% of the 
data have been taken (and in below average seeing conditions).

\section{\label{release}Summary and data release}
The Gemini Frontier Fields Campaign complements \textit{HST} observations redwards of $1.6\mu$m for
three galaxy clusters, MACS J0416.1-2403, Abell 2744 and Abell 370. Using GeMS/GSAOI, the first MCAO 
system in use at an 8m telescope, near diffraction-limited images on angular scales larger than 
$1^{\prime}$ are obtained in \ks-band. We make the fully calibrated co-added images and weights publicly 
available\footnote{{\tt http://www.gemini.edu/node/12254/}}. This paper describes the observations 
and data reduction of the first two clusters observed, MACS J0416.1-2403 (complete) and Abell 2744 
(incomplete). Abell 370 is scheduled for observation in 2015.

We release different co-added images resampled to 0\farcs02, 0\farcs03 and 0\farcs06 
pixel$^{-1}$ (high, medium and low resolution stacks). The first preserves the native pixel 
scale of GSAOI, whereas the other two use identical plate scales and image geometries as 
the \textit{HST} data release (their v1.0).

\begin{figure}[t]
  \includegraphics[width=1.0\hsize]{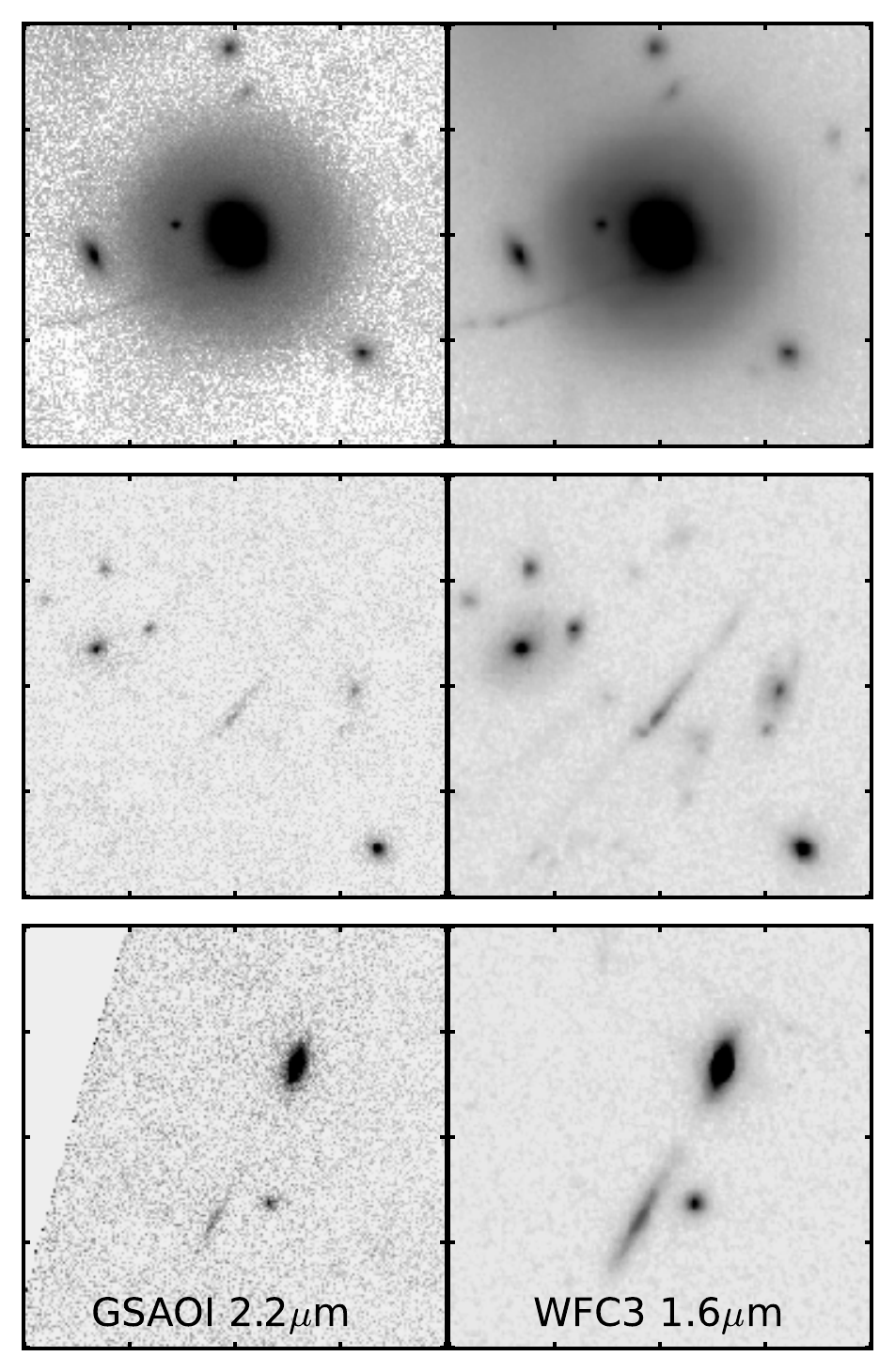}
  \caption{\label{macs0416_cutout}{Comparison of strongly lensed features in MACS J0416.1-2403 seen by 
      GSAOI and \textit{HST/WFC3}. The areas measure $12^{\prime\prime}\times12^{\prime\prime}$, and have 
      North up and East left.}}
\end{figure}

Natural seeing conditions have varied during the observing nights for MACS J0416.1-2403. We 
provide two different versions for the high and medium resolution stacks, optimized for seeing 
($35\%$ of all exposures) and depth ($85\%$). For the low resolution stack only one deep version is 
provided comprised of all usable exposures, as PSF anisotropies and softer seeing become insignificant. 
For Abell 2744 all currently available exposures were used. Table \ref{datarelease} 
summarizes the key properties of the stacked images.

\begin{figure*}[t]
  \includegraphics[width=1.0\hsize]{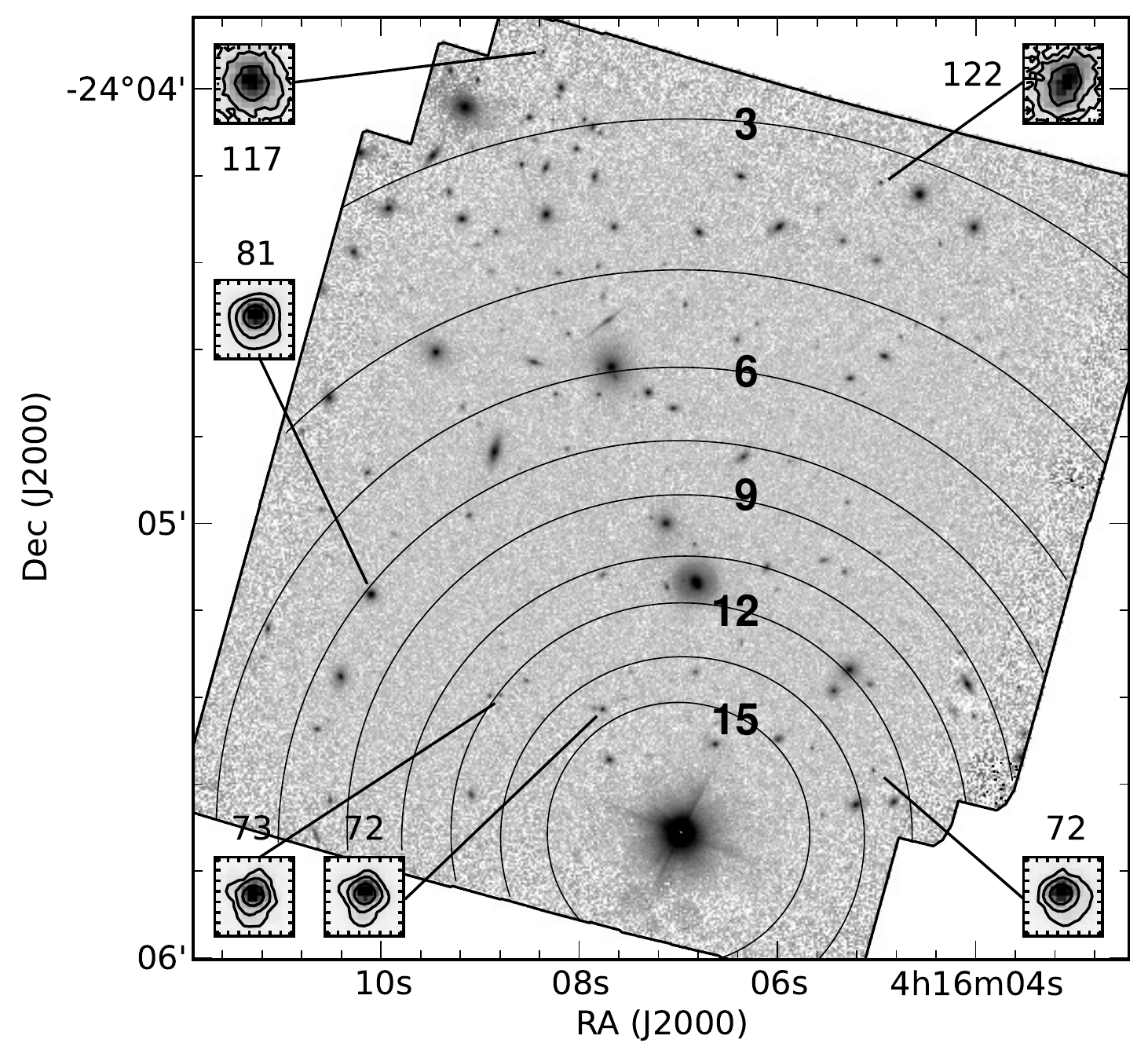}
  \caption{\label{macs0416_coadd}{PSF of the 6 brightest field stars in our medium resolution 
      \textit{good seeing} stack. The extent of the small boxes is 0\farcs3$\times$0\farcs3, and the 
      contour levels therein correspond to 0.1, 0.2 and 0.5 times the stars' peak fluxes. The PSF is 
      fairly uniform and increases with growing distance from the single natural guide star (bright 
      source at the bottom). Only the upper right star displays significant ellipticity. The numeric 
      values next to each box correspond to the direct FWHM in milli-arcseconds, measured with 
      {\tt IRAF/imexam}. The large concentric contours around the natural guide star show the 
      expected Strehl ratio for the given guide configuration, increasing from 3\% to 15\%.
      The expected average Strehl ratio is 6.9\%, corresponding to a FWHM of 90 mas, consistent with 
      the delivered image quality. For reference, the diffraction limited FWHM for GeMS/GSAOI in 
      \ks-band is 55 milli-arcseconds \citep{nrv14}.}}
\end{figure*}

Only one NGS is available for each of the galaxy clusters and GeMS/GSAOI. This results 
in increasing PSF variations and FWHM as a function of separation from the NGS 
(Fig. \ref{macs0416_coadd}). The performance variation over the field is predicted well based 
on numerical simulations, which are available in the Gemini Observing 
Tool\footnote{{\tt http://www.gemini.edu/sciops/instruments/gems/ 
gems-performance?q=node/11864\#Asterism\_visualization}}. With a FWHM of 0\farcs07$-$0\farcs10 
we improve upon \textit{HST/WFC3}'s angular resolution by a factor of two, albeit over a smaller field 
of view ($100^{\prime\prime}\times110^{\prime\prime}$). We reach a $5\sigma$ depth for extended 
sources of $K^{\rm lim}_s=25.6$ AB mag. This demonstrates that MCAO at Gemini South works well 
even for high galactic latitude fields where natural guide stars are scarce, opening 
a new window onto the distant Universe. We have also shown that current data reduction techniques, 
developed for classical imaging, are well suitable for this type of MCAO data.

\subsection{Future work}
In 2015 and 2016 the MCAO system at Gemini Observatory will receive two very significant upgrades.
First, new NGS wavefront sensors enable us to guide on $V\sim17.5$ mag stars (currently $V\sim15$ mag), which 
will dramatically increase the sky coverage, and the number of targets for which 
triangular guide star constellations are available. Second, funding has been secured for the purchase 
of a Toptica \citep{kpc11} laser offering increased LGS brightness, easier maintenance and operation, and
lower losses due to simplified beam transfer optics. 

\begin{acknowledgements}
\section*{Acknowledgments}
We thank Nancy Levenson and Markus Kissler-Patig for granting 
us directors' discretionary time, and the enonymous referee for the valuable comments that 
helped us improve our paper. This work utilizes gravitational 
lensing models produced by PIs Bradač, Ebeling, Merten \& Zitrin, Sharon, and Williams 
funded as part of the \textit{HST} Frontier Fields program conducted by STScI. STScI is operated 
by the Association of Universities for Research in Astronomy, Inc. under NASA contract 
NAS 5-26555. The lens models were obtained from the Mikulski Archive for Space Telescopes 
(MAST). The astrometric calibrations of the GSAOI data have been based on secondary reference catalogs
derived from \textit{VLT/HAWK-I} observations, made with ESO Telescopes at the La Silla Paranal 
Observatory under program ID 092.A-0472.
\end{acknowledgements}

\bibliography{mybib}

\begin{appendix}
\begin{figure*}[t]
  \includegraphics[width=1.0\hsize]{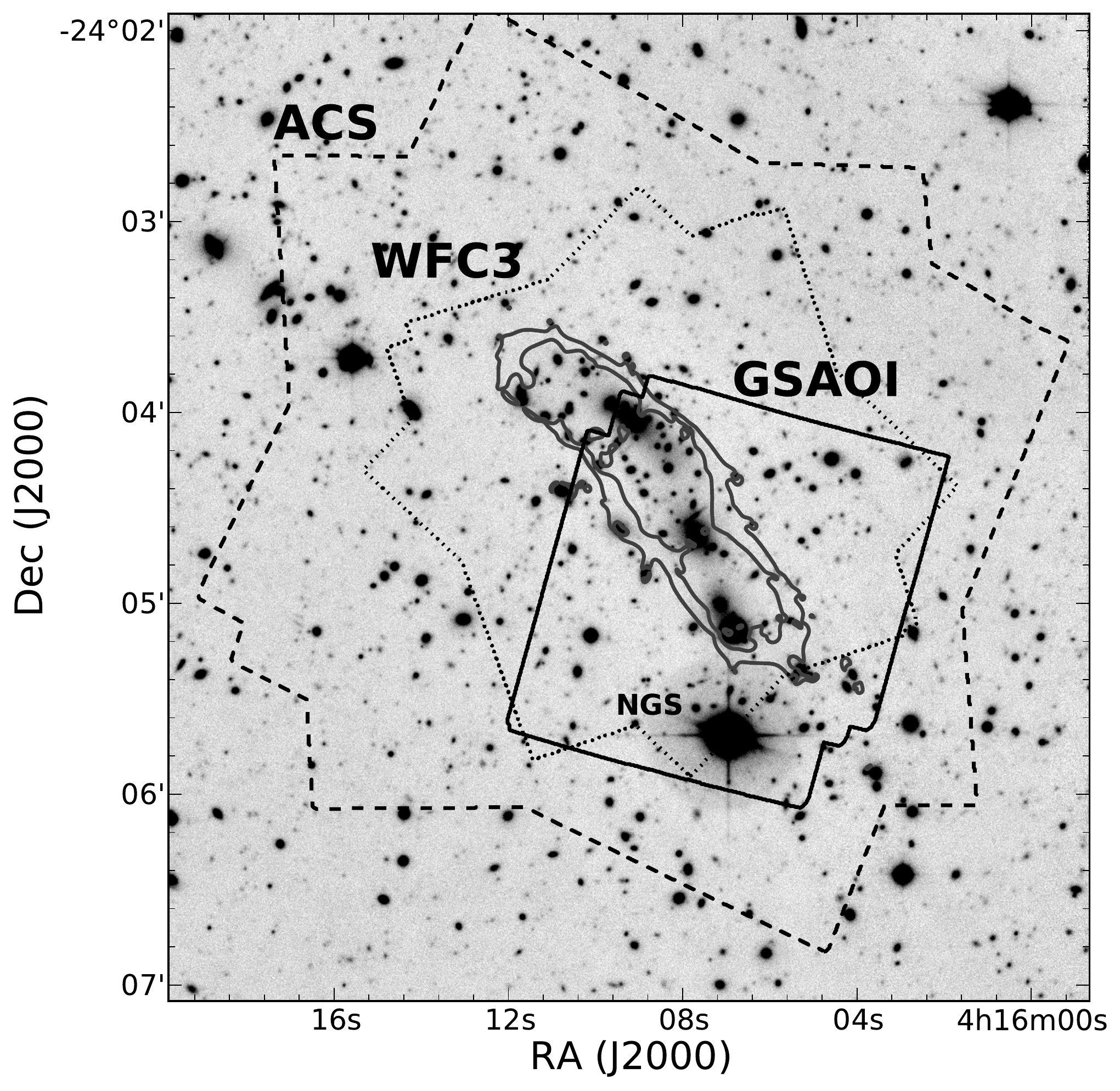}
  \caption{\label{macs0416_layout}{Layout of the \textit{HST/ACS}, \textit{HST/WFC3} and GSAOI fields for MACS 
      J0416.1-2403. The (smoothed) gray contours indicate the area where a magnification higher than 20
      is expected for a $z=9$ source \citep[from][]{rjl14}. The NGS has been labeled, too.}}
\end{figure*}

\begin{figure*}[t]
  \includegraphics[width=1.0\hsize]{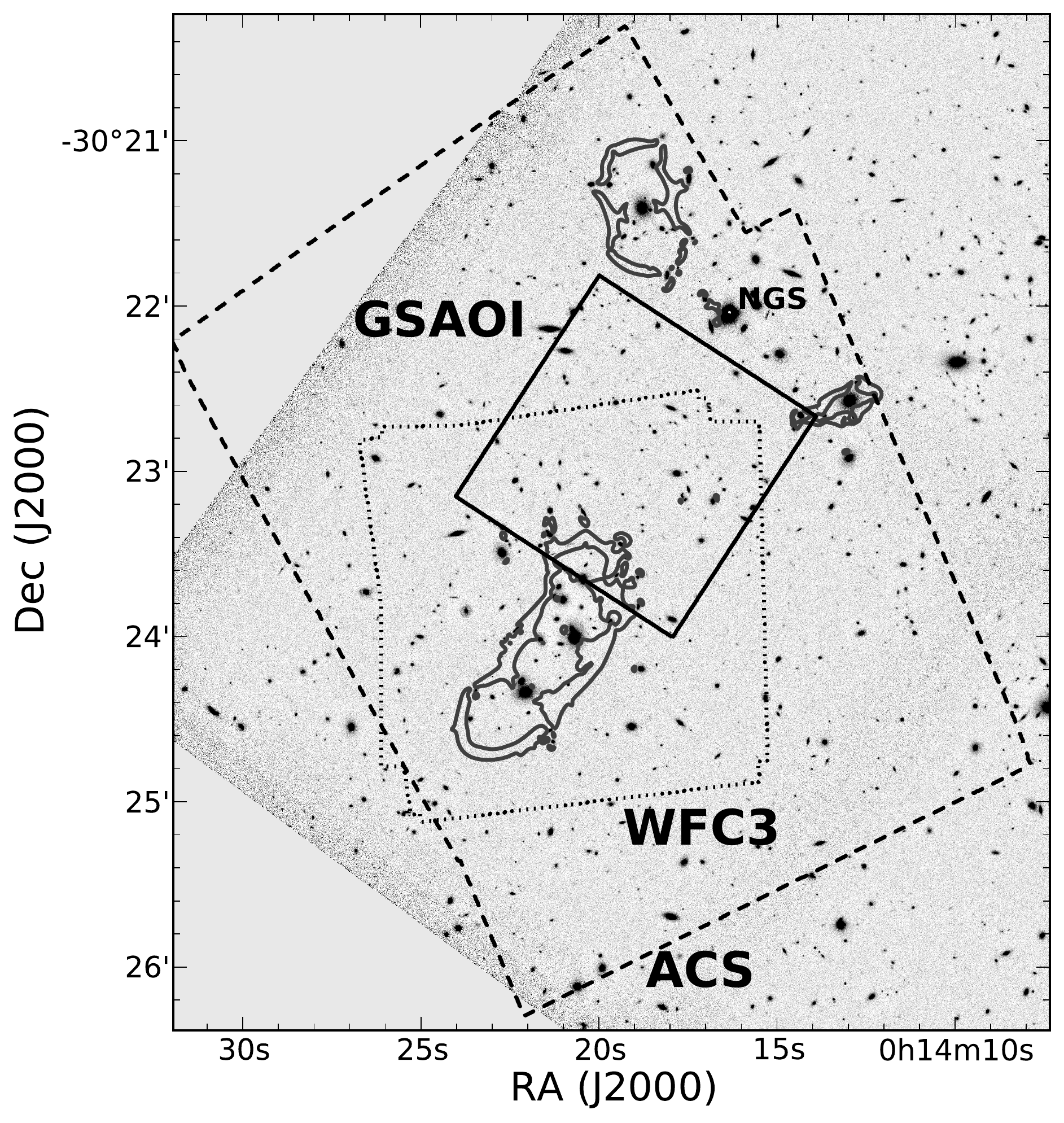}
  \caption{\label{a2744_layout}{Same as Fig. \ref{macs0416_layout}, for Abell 2744. Note that the NGS is too far
  away from the cluster center for GSAOI to cover the area with highest magnification.}}
\end{figure*}

\begin{figure*}[t]
  \includegraphics[width=1.0\hsize]{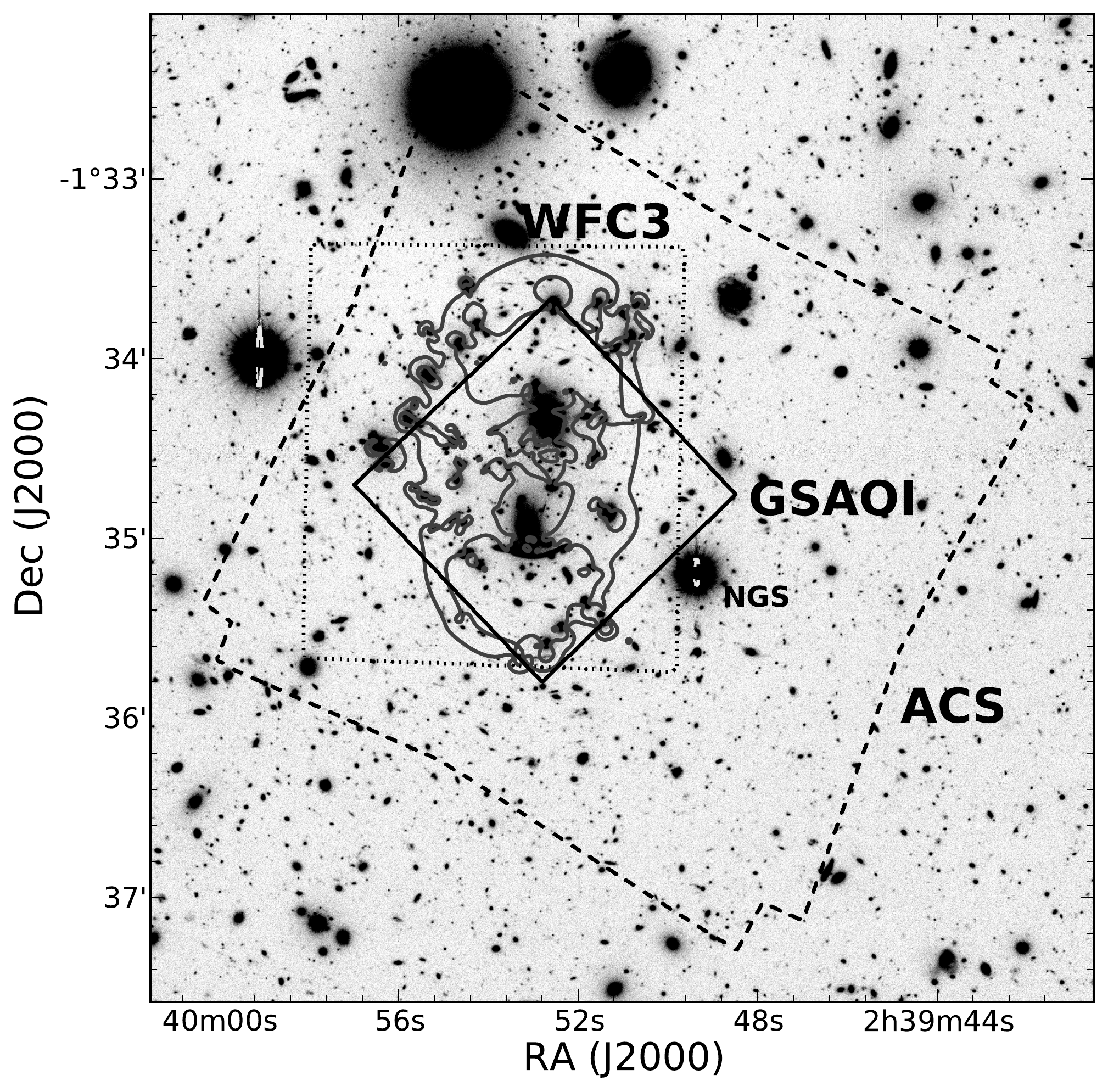}
  \caption{\label{a370_layout}{Same as Fig. \ref{macs0416_layout}, for Abell 370. Note that the \textit{HST} 
  layouts are subject to change as further imaging is pending. Likewise, the final GSAOI pointing has not 
  yet been chosen.}}
\end{figure*}

\end{appendix}

\end{document}